\documentclass[12pt]{iopart}
\pdfoutput=1
\usepackage{iopams}

\usepackage{amsopn}   %instead use this
\usepackage{amssymb}
\usepackage{epsfig}
\usepackage[T1]{fontenc}
\usepackage[utf8]{inputenc}
\usepackage{color}
\usepackage[normalem]{ulem}
\usepackage{tikz}
\usetikzlibrary{positioning}
\usepackage{tabularx}
\usepackage{enumitem}
\usepackage{graphicx}
\usepackage{caption}
\usepackage{subcaption}
\usepackage{ragged2e}
\DeclareCaptionJustification{justified}{\justifying}
\usepackage{dcolumn}% Align table columns on decimal point
\usepackage{bm}% bold math
\usepackage{comment}
\usepackage{siunitx}
\usepackage[title]{appendix}
\usepackage{booktabs}
\usepackage{float}
\usepackage{multirow}
\usepackage{lipsum}
\usepackage{tabulary,xcolor}
\usepackage{amsfonts}
\usepackage{verbatim} 
\usepackage{url,morefloats,floatflt,cancel}
\usepackage{hyperref}
\usepackage{enumitem}
\def\d{\mathrm{d}}

\begin{document}
\title[Universal distributions in the Blume-Capel and Baxter-Wu models]{Universal energy and magnetisation distributions in the Blume-Capel and Baxter-Wu models}

\author{A.~R.~S. Mac\^edo\textsuperscript{1,2},
J.~A. Plascak\textsuperscript{1,3,4$\star$},
A. Vasilopoulos\textsuperscript{5},
N.~G. Fytas\textsuperscript{5$\dagger$},
M. Akritidis\textsuperscript{6},
and
M. Weigel\textsuperscript{7$\ddagger$}}

\address{$^1$Departamento de F\'\i sica, Instituto de Ci\^encias Exatas, Universidade Federal de Minas 
Gerais, C.P. 702. 30123-970 Belo Horizonte, MG - Brazil}

\address{$^2$Instituto Federal do Maranh\~ao - Campus Imperatriz,  65919-050, Imperatriz, MA - Brazil}

\address{$^3$Universidade Federal da Para\'iba, Centro de Ci\^encias Exatas e da Natureza - Campus I, 
Departamento de F\'isica - CCEN Cidade Universit\'aria 58051-970 Jo\~ao Pessoa, PB - Brazil}

\address{$^4$Center for Simulational Physics, University of Georgia, 30602 Athens, GA - USA }

\address{$^5$School of Mathematics, Statistics and Actuarial Science, University of Essex, Colchester CO4 3SQ, United Kingdom}

\address{$^6$School of Computing, Mathematics and Data Science, Coventry University, Coventry CV1 5FB, United Kingdom}

\address{$^7$Institut für Physik, Technische Universität Chemnitz, 09107 Chemnitz, Germany}

\begin{center}
    $\star$ \href{mailto:pla@uga.edu}{\small pla@uga.edu}\,,\quad
    $\dagger$ \href{mailto:nikolaos.fytas@essex.ac.uk}{\small nikolaos.fytas@essex.ac.uk}
    $\ddagger$ \href{mailto:martin.weigel@physik.tu-chemnitz.de}{\small martin.weigel@physik.tu-chemnitz.de}
\end{center}

\date{\today}

\begin{abstract}
We analyse the probability distribution functions of the energy and magnetisation of the two-dimensional Blume-Capel and Baxter-Wu models with spin values $S \in \{1/2,1, 3/2\}$ in the presence of a crystal field $\Delta$. By employing extensive single-spin flip Monte Carlo simulations and a recently developed method of studying the zeros of the energy probability distribution we are able to probe, with a good numerical accuracy, several critical characteristics of the transitions. Additionally, the universal aspects of these transitions are scrutinised by computing the corresponding probability distribution functions. The energy distribution has been underutilised in the literature when compared to that of the magnetisation. Somewhat surprisingly, however, the former appears to be more robust in characterising the universality class for both models upon varying the crystal field $\Delta$ than the latter. Finally, our analysis suggests that in contrast to the Blume-Capel ferromagnet, the Baxter-Wu model appears to suffer from strong finite-size effects, especially upon increasing $\Delta$ and $S$, that obscure the application of traditional finite-size scaling approaches.
\end{abstract}

\submitto{Journal of Statistical Mechanics}

\maketitle

\section{Introduction}
\label{sec:intro}

The Blume-Capel (BC) model is defined by a spin-$1$ Ising Hamiltonian with a single-ion uniaxial crystal-field anisotropy~\cite{blume,capel}. The fact that it has been very widely studied in statistical and condensed-matter physics is explained not only by its relative simplicity and the fundamental theoretical interest arising from the richness of its phase diagram, but also by a number of different physical realisations of variants of the model, ranging from multi-component fluids to ternary alloys and $^3$He-$^4$He mixtures~\cite{lawrie}. The zero-field model is described by the Hamiltonian 
\begin{equation}
{\cal{H}}^{\rm (BC)}=-J\sum_{\langle ij\rangle}{\sigma_i}{\sigma_j}+\Delta\sum_i \sigma_i^2, 
\label{bcm}
\end{equation}
where $J>0$ is a ferromagnetic exchange interaction and $\Delta$ denotes the crystal-field coupling. The first sum is over all nearest-neighbours $\langle ij\rangle$ and the second over all spins of the lattice. Our numerical work is focused on the square lattice but the model is, of course, more general. For $S=1$ the variables $\sigma_i$ take on the values $\sigma_i=0,\pm1$, while for general $S = 1$, $3/2$, $2$, $\ldots$ one has $\sigma_i\in\{-S,-S+1,\ldots,S-1,S\}$  --- importantly, for integer $S$, this includes $\sigma_i = 0$, while for half-integer $S$ it does not.

The Blume-Capel model on different lattice geometries and for various $S$ values has been studied extensively over the years. Methodologically, a vast variety of approximation methods have been used to tackle the problem, such as mean-field theory~\cite{blume,capel,pla0,ricardo}, renormalisation group~\cite{moss}, finite-size scaling and conformal field theory~\cite{jurgen,beale,xav,dias17,jung}, Monte Carlo simulations~\cite{jain,nigel,pla1,fytas1,fytas2,kwak,fytas3}, and series expansions~\cite{butera} (see also references therein). Although most of the simulation studies mentioned above have focused on $S=1$ or $S=3/2$, a number of general features of the phase diagram in the $(\Delta-T)$ plane follow from predictions of mean-field theory. In particular~\cite{pla0,ricardo}:
\begin{enumerate}[topsep=3pt,itemsep=1ex]
\item For integer values of $S$ a second-order transition line with a decreasing critical temperature $T_{\rm c}(\Delta)$ as $\Delta$ increases meets a first-order transition line at a tricritical point. This first-order transition line reaches the point of zero temperature ($T=0$) at $\Delta_0=zJ/2$, where $z$ is the coordination number of the lattice. From the point $T=0$ and $\Delta = \Delta_0$, $S-1$ additional first-order transition lines emerge as the temperature rises and all end up at independent double critical endpoints~\cite{pla0}.
\item  For half-integer $S$ values, the second-order transition line extends to all values of the crystal field. However, from $T=0$ and $\Delta=\Delta_0$, $S-1/2$ additional first-order transition lines emerge now as the temperature rises, all ending up at independent double critical endpoints located below the critical transition line~\cite{pla0}.
\item The critical lines as well as the double critical endpoints are in the same universality class as the regular Ising ferromagnet, regardless of the particular value of $S$.
\end{enumerate}
In addition, for the $S=1$ square-lattice model the location of the tricritical point is known with a high numerical accuracy to lie at $(\Delta_{\rm t}, T_{\rm t}) \approx (1.9660(1), 0.6080(1))$~\cite{kwak}.

On the other hand, the Baxter-Wu (BW) model was first introduced by Wood and Griffiths~\cite{wood} as a system which does not exhibit invariance under a global inversion of all spins. The Hamiltonian of the model, again augmented by a crystal-field coupling term, reads
\begin{equation}
{\cal{H}}^{\rm (BW)}=-J\sum_{\langle ijk\rangle}{\sigma_i}{\sigma_j}\sigma_k +\Delta\sum_i \sigma_i^2,
\label{bwh}
\end{equation}
where, as in equation~(\ref{bcm}), $J > 0$, but now the first sum extends over all elementary triangles $\langle ijk\rangle$ of the triangular lattice. In the original model of reference~\cite{wood} $\sigma_i = \pm 1/2$ (then without crystal-field term, i.e., for $\Delta = 0$), while the spin-$1$ case has $\sigma_i = 0,\pm 1$. It is easily seen that the presence of three-spin interactions results in a four-fold degeneracy of the ground state:  there is one ferromagnetic state with all spins up, and three ferrimagnetic states with down-spins in two sublattices and up-spins in the third. A useful representation of the sublattice structure can be found in figure 1 of reference~\cite{vasilopoulos22}. 

An exact solution of the original $S=1/2$ Baxter-Wu model was provided early on by Baxter and Wu~\cite{baxter73,baxter_book}, supplying the critical exponents $\alpha = 2/3$, $\nu = 2/3$, and $\gamma = 7/6$. In the following, it was also shown that its critical behaviour corresponds to a conformal field theory with central charge $c = 1$~\cite{alcaraz97,alcaraz99}. Due to the four-fold symmetry of the ground state it is expected that the critical behaviour of the Baxter-Wu model is in the same universality class as the $q = 4$ Potts model in two dimensions~\cite{domany78}\footnote{As pointed out by E. Domany (private communication) this fact was first noticed by R.B. Griffiths.}. While, therefore, the critical exponents of the two models are identical, the same does not apply to the scaling corrections: the $4$-state Potts model exhibits logarithmic corrections with the system size~\cite{wu82}, whereas the Baxter-Wu model has power-law corrections with a predicted correction-to-scaling exponent $\omega = 2$~\cite{barber76}.  Other models in the same universality class also exist. Examples are the square-lattice Ising model with mixed two- and three-spin interactions~\cite{turban82} as well as a quantum version of the model~\cite{igloi87}.

Compared to the Blume-Capel model, the phase diagram of the Baxter-Wu model in the presence of a crystal field is not so well understood, not only regarding the presence of multicritical points but also regarding the question of universality along the second-order transition line. Based on an analogy between the BW model and a diluted $4$-state Potts model,  Nienhuis \emph{et al.}~\cite{nien} pointed out that the phase diagram of the $S=1$ case exhibits a line of continuous transitions as well as a regime of first-order transitions connected through a multicritical point. In fact, this corresponds to a tetracritical line joining a quintuple line (coexistence of five phases) at a pentacritical point (see, for example, reference~\cite{pla3} for the terminology relating to multiple and multicritical points). Conversely, Kinzel \emph{et al.}~\cite{kinzel}, using a finite-size scaling method, conjectured that a continuous transition would only occur at $\Delta\rightarrow-\infty$ (the pure Baxter-Wu model). More recent works using Monte Carlo simulations and conformal invariance have indeed favoured the existence of a pentacritical point at finite values of $\Delta$ for spin $S=1$~\cite{dias17,maria,jorge,maria2,maria3}. With this observation the phase diagram of the $S=1$ Baxter-Wu model turns out to be rather analogous to that of the Blume-Capel model for the same value of spin. Nevertheless, an accurate estimate of the location of the pentacritical point (pp) is currently not available, complicating the analysis of numerical data in the area of this putative multicritical point: Note the discrepancy in the estimates by Dias \emph{et al.}~\cite{dias17}, $(\Delta_{\rm pp}, T_{\rm pp}) \approx (0.8902, 1.4)$, and Jorge \emph{et al.}~\cite{jorge}, $(\Delta_{\rm pp}, T_{\rm pp}) \approx (1.68288(62), 0.98030(10))$; see also figure 2 in reference~\cite{arilton}. Recently, there have even been arguments in favour of a first-order transition for the spin $S=1/2$ model~\cite{lucas1st} (a mean-field treatment of the $S = 1/2$ model~\cite{moallison} erroneously predicts a first-order transition).

Although it has received little attention to date, it is of course also possible to study the model (\ref{bwh}) for spin $S > 1$. For $S=3/2$, to the best of our knowledge, the only known result stems from a finite-size scaling and conformal invariance study~\cite{dias17} which suggested that one has, along the second-order line (again, a tetracritical line), and close to the region where $\Delta/J\sim z/2$, a short segment of a first-order line with five coexisting phases, \emph{i.e.}, a quintuple line. To the left of this quintuple line one has a pentacritical point and to the right a tetracritical endpoint. From this tetracritical endpoint a low-temperature octuple line (eight coexisting phases) goes down to the point $T=0$ and $\Delta/J = 3.25$~\cite{dias17}. Thus, the phase diagram of the Baxter-Wu model with $S=3/2$ appears to differ significantly from that of the corresponding Blume-Capel model.  
The universality class of the second-order transition line has also been studied for both values of the spin, $S=1$ and $S=3/2$. From renormalisation-group arguments one expects the second-order transition line to remain in the universality class of the $S=1/2$ model and, therefore, in that of the $4$-state Potts model. Indeed, this has been clearly shown to be the case by comparing critical exponents and other renormalisation-group invariants for both spin $S=1$~\cite{dias17,vasilopoulos22,arilton} and $S=3/2$~\cite{dias17} models for a wide range of crystal-field values in the regime $\Delta < 0$. Nonetheless, for positive values of $\Delta$ and in closer proximity to the multicritical point, the model develops strong finite-size effects and numerical estimates of critical quantities show a systematic shift away from the expected results, as has been recently reported in reference~\cite{arilton} for the $S=1$ model. Even for the case with $\Delta = 0$, carefully crafted simulations utilising rather large system sizes ($L = 240$, where $L$ defines the linear dimension of the lattice) were needed for a clear demonstration of universality in this regime~\cite{arilton}. For $\Delta > 0$, such effects are expected to be even stronger. While it is of course conceivable that the model transitions to a new universality class for $\Delta > 0$, we believe that this is rather unlikely given that there is no change in symmetry. In this respect, the analysis of universal probability distribution functions (PDFs) to be discussed below is hoped to shed new light on this controversial aspect of the problem.

In the present work we focus on issues relating to universality in both models along the second-order transition lines in the $\Delta-T$ plane, studying them in depth by considering the numerically accessible PDFs. The majority of our Monte Carlo simulations are performed via the single spin-flip Metropolis algorithm complemented by histogram methods and finite-size scaling arguments~\cite{metro,landau}. In particular, for dedicated values of the crystal field we locate the critical points and compute the critical exponent $\nu$ of the correlation length by means of a recently developed technique revolving around the zeros of the energy probability distribution~\cite{bis1,bis2,bis3}. We then consider the corresponding critical distributions of the energies and magnetisations that are expected to be universal and can hence serve as sensitive indicators for determining universality classes. We remind the reader here that, for the Baxter-Wu model the energy and magnetisation PDFs have previously been considered for the particular case $S=1/2$~\cite{marti,adler,velo0,velo}.

The rest of this paper is organised as follows: In section~\ref{sec:theory} we elaborate on the necessary theoretical background, namely on the method of energy probability distribution zeros and the pathway to the universal energy and magnetisation probability distributions. Subsequently, in section~\ref{sec:simulations}, we provide an outline of the employed Monte Carlo methods and simulation protocols. Section~\ref{sec:results} contains the presentation and critical discussion of the numerical results for both Blume-Capel and Baxter-Wu models. Finally, the paper concludes with a summary and some additional remarks in section~\ref{sec:summary}. 

\section{Theoretical background}
\label{sec:theory}
Our study of the Blume-Capel and Baxter-Wu models is twofold: an analysis of the zeros of the energy probability distribution (EPD) yields our main estimate of the transition temperatures and the shift exponent $1/\nu$. Motivated by the observations made there, the question of universality is then examined in detail by a comparison of the universal critical distributions of the energy and magnetisation of the models. Here we provide the necessary background for these studies.

\subsection{Energy probability distribution zeros}
\label{zeros}

Consider a system of statistical mechanics with a discrete energy spectrum with levels $E=E_n=\varepsilon_0+n\varepsilon$, $n=0,1,\ldots,\mathcal{N}$, with $\varepsilon_0$ being the ground state energy and $\varepsilon$ the level spacing. We can write the partition function of this system in the following form:
\begin{equation}
{\cal Z}_\beta=\sum_Eg(E)e^{-\beta E}=e^{-\beta \varepsilon_0}\sum_{n=0}^{\cal N}
g_ne^{-\beta n\varepsilon}= e^{-\beta \varepsilon_0}\sum_{n=0}^{\cal N}g_nz^n,
\label{zn}
\end{equation}
where $g(E)$ is the number of states with energy $E$ (\emph{i.e.}, the density of states) and $\beta=1/(k_\mathrm{B}T)$ with $k_\mathrm{B}$ being Boltzmann's constant and $T$ the temperature; in the two last identities we use the shorthand notation $g_n=g(E_n)$ and $z=e^{-\beta \varepsilon}$. 

As noted by Fisher~\cite{fisher}, the analytic structure of the partition function and hence the occurrence of phase transitions can be understood from its factorised form that follows immediately once one knows all of its \emph{zeros}. For a finite system $\mathcal{Z}$ is a polynomial of degree $\mathcal{N}$ in $z$ and it hence has exactly $\mathcal{N}$ complex zeros. Since $g_n$ is real and non-negative, none of the zeros are real, but they come in complex conjugate pairs. However, as shown by Fisher, for $\mathcal{N}\to\infty$ at least one root approaches the real axis at $z_c=e^{-\beta_\mathrm{c} \varepsilon}$, and this event indicates the occurrence of a phase transition at $\beta_\mathrm{c} = 1/k_\mathrm{B} T_\mathrm{c}$. For finite lattices one cannot directly investigate this limit $\mathcal{N}\to\infty$ but, instead, one usually studies the \emph{dominant} zero, \emph{i.e.}, the zero closest to the real axis. Since it is not completely straightforward to sample partition function zeros in a Monte Carlo study, a slightly modified approach was proposed in reference~\cite{bis1}. Inserting unity in the form  $1=e^{-\beta_0 E}e^{+\beta_0 E}$ into equation~(\ref{zn}), where $\beta_0$ is some reference inverse temperature, we immediately obtain the expression
\begin{equation}
{\cal Z}_{\beta} = e^{-\Delta\beta \varepsilon_0}\sum_{n=0}^{\cal N}h_{\beta_0}(n)x^n,
\label{zb}
\end{equation}
where $\Delta\beta=\beta-\beta_0$, $h_{\beta_0}(n)=g_n e^{-\beta_0 E}$, and $x=e^{-\Delta\beta\varepsilon}$. In this way, the set $\{x_i\}$ of zeros of the above equation are just the renormalised set of Fisher zeros $\{z_i\}$ resulting from equation~(\ref{zn}), since $x=e^{-\beta \varepsilon}/e^{-\beta_0 \varepsilon}=z/e^{-\beta_0 \varepsilon}$. We note that $h_{\beta_0}(n)=g_n e^{-\beta_0 E}$ is the (unnormalised) energy probability distribution at inverse temperature $\beta_0$. Hence it can be easily estimated numerically from an energy histogram sampled at $\beta_0$. Computing the zeros of an estimate of equation (\ref{zb}) with $h_{\beta_0}(n)$ replaced by the histogram $\hat{h}_{\beta_0}(n)$ hence provides an easy pathway towards an analysis of the partition function zeros. Since the tails of the EPD are less populated, a cutoff $\delta$ can be introduced by neglecting all coefficients $h_{\beta_0}(n)\le\delta$, reducing significantly the degree of the polynomial for larger lattices. In order to avoid having to cope with the adverse numerical implications of very large polynomial coefficients, one can further normalise the histogram by considering $h_{\beta_0}(n)/h_{\beta_0}^{(\max)}$, where $h_{\beta_0}^{(\max)}$ is the maximum value of the histogram.

Now, when $\beta_0=\beta_\mathrm{c}$, the zero corresponding to the phase transition for the infinite system appears at $x_\mathrm{c}=(1,0)$. For a finite lattice of size $L$ this implies that the dominant zero $x_{\rm c}^{L}$ should be close to the point $(1,0)$ if we choose $\beta_0 \approx \beta_\mathrm{c}^L$. This observation suggests an iterative approach for locating $\beta_\mathrm{c}^L$ by considering the location of the dominant zero $x_\mathrm{c}^L$ for a given choice of $\beta_0$~\cite{bis1,bis2,bis3}. One starts with an initial guess for $\beta_\mathrm{c}^L$ denoted by $\beta_0^{j=0}$ and iterates through the following steps:
\begin{enumerate}
\item[(1)] Simulate the system at $\beta = \beta_0^j$ and construct a histogram $\hat{h}_{\beta_0^j}$.
\item[(2)] Determine all the zeros $x_i^j$, $i=1,\ldots,\mathcal{N}$ of the polynomial
\[
\sum_{n=0}^\mathcal{N} \hat{h}_{\beta_0^j}(n) x^n.
\]
\item[(3)] Find the dominant zero $x_{\rm c}^j$. Then:
\begin{itemize}
\item[(a)] if $x_\mathrm{c}^j$ is (to a prescribed level of accuracy) close enough to the point $(1, 0)$, take $x_\mathrm{c}^L=x_\mathrm{c}^j$ and stop;
\item[(b)] else, take 
\begin{equation}
\beta_0^{j+1} = -\varepsilon^{-1}\ln{ [\operatorname{Re}(x_\mathrm{c}^j)]} + \beta_0^j
\label{iter}
\end{equation}
and repeat at step (1). 
\end{itemize}
\end{enumerate}
In the above equation~(\ref{iter}), $\operatorname{Re} (x_{\rm c}^j)$ denotes the real part of the complex root $x_{\rm c}^j$. At the end of this process, not only do we have $x_{\rm c}^L$ but also $\beta_0^j=\beta_0^L \approx \beta_c^L$, the desired pseudocritical inverse temperature, corresponding to the temperature $T_{\rm c}^L$ of the most relevant zero for the lattice size $L$. We can then make use of the established finite-size scaling form for pseudocritical inverse temperatures,
\begin{equation}
T_{\rm c}^L=T_{\rm c}+bL^{-1/\nu}\left(1+b^\prime L^{-\omega}\right),
\label{tc}
\end{equation}
where $T_{\rm c}$ is the critical temperature of the infinite system, $b$ and $b^\prime$ are non-universal constants, $\nu$ the critical exponent of the correlation length, and $\omega$ the correction-to-scaling exponent. An analogous behaviour is expected for the scaling of the dominant zero~\cite{bis1}
\begin{equation}
\label{eq:x_c}
x_{\rm c}^L=x_{\rm c}+bL^{-1/\nu}(1+b'L^{-\omega}).
\end{equation}
Thus, the real part approaches $\operatorname{Re}  (x_c^L) \to 1$, while the imaginary part goes to zero as~\cite{bis1}
\begin{equation}
\operatorname{Im}(x_{\rm c}^L) \sim L^{-1/\nu}(1+b'L^{-\omega}).
\label{imx}
\end{equation}
Although in the above scaling equations~(\ref{eq:x_c}) and (\ref{imx}) we have included a correction-to-scaling term proportional to $L^{-\omega}$, it turns out that in our numerical data these corrections are very small and thus the extra term was not included in those fits.

\subsection{Critical energy and magnetisation probability distribution functions}
\label{upd}

Universality classes are most often characterised by the values of critical exponents, sometimes also by comparison of universal amplitude ratios. Much less attention is being paid to universal \emph{distribution functions} of the extensive thermodynamic variables such as the energy and magnetisation~\cite{binder,milchev}, although they naturally contain much more information than the single numbers of exponents and amplitude ratios. Here, we find that such distributions are quite useful as they indeed provide a fingerprint of the underlying universality class. Most of the first studies on the topic involved the computation of the magnetisation (or order-parameter) PDFs in Ising-like models~\cite{bruce,binder,wild1,wild2,wild3,wild4,rev}, due to the main interest being primarily the location of first-order transition lines and multicritical points. Here, we will also consider the energy PDF, and find it rather more useful for comparing models than the magnetisation distributions. In the present section, we will describe the formalism for both cases in a unified language.

According to fundamental scaling arguments, the general form of the critical PDF of the density $x=X/N$ of an extensive thermodynamic variable $X$ in a system of edge length $L$ and number of sites $N=L^d$ is expected to take the form~\cite{binder,milchev}
\begin{equation}
P_L(x) = aL^uP^*[bL^v(x-x_0)] = aL^uP^*[bL^v\chi],
\label{pdsr}
\end{equation}
where $x_0 = X_0/N = \langle X\rangle/N$ is the expectation value of $x$, $u$ and $v$ are critical exponents, and $a$ and $b$ are non-universal metric constants. Importantly, here $P^*$ is the \emph{universal} scaling function related to the PDF of $x$. Normalisation of $P_L(x)$ as a probability distribution implies that $u = v$ and one can assume without loss of generality that $a=b$~\cite{binder}. It is often convenient to consider directly the shifted scaling variable $\chi = x-x_0$ as we shall do below.

In a numerical setting, we can generate an estimate of $P_L$ by constructing a histogram $\hat{P}_L$ of values $X$ from a time series $X_t$, $t=1,\ldots, T$ of length $T$ of measurements of $X$ taken at (or very close to) the critical temperature. We assume here that $X$ has a discrete spectrum with $X \in \Omega_X = \{X_\mathrm{min}, X_\mathrm{min}+\Delta X, X_\mathrm{min}+2\Delta X, \ldots, X_\mathrm{max}\}$ (for a continuous spectrum one also naturally arrives at this form through some binning procedure). Formally, the normalised histogram is then given by
\begin{equation}
  \hat{P}_L(X) = \frac{1}{T}\sum_{t=1}^T \delta_{X_t,X},\;\;\;X\in\Omega_X.
  \label{dp}
\end{equation}
By construction, one has $\sum_{X\in\Omega_X} \hat{P}_L(X) = 1$ and we can use the empirical histogram to estimate $X_0$
\[
\hat{X}_0 = \frac{1}{T}\sum_{t=1}^T X_i = \sum_{X\in\Omega_X} \hat{P}_L(X) X.
\]
Since $X$ is extensive, the range $X_\mathrm{max}-X_\mathrm{min}$ grows proportional to $N$, while we can assume that the spacing $\Delta X$ is independent of $N$. Then, the possible values for the density $x=X/N$ become quasi continuous for large $N$ since we can write
\begin{equation}
    1 = \sum_{X\in\Omega_X} \hat{P}_L(X) = \sum_{X=X_\mathrm{min}}^{X_\mathrm{max}} \frac{\hat{P}_L(X)}{\Delta X/\omega_X} \frac{\Delta X}{\omega_X}
    \stackrel{N\gg 1}{\longrightarrow} \int_{x_\mathrm{min}}^{x_\mathrm{max}} \frac{\hat{P}_L(Nx)}{\Delta X/\omega_X}\d x,
\end{equation}
where $\omega_X = |\Omega_X|$ denotes the number of discrete values of the variable $X$; clearly, for $N\gg 1$ one finds that $\mathrm{d}x = \Delta X/\omega_X \ll 1$ and $x$ approaches a continuous variable. We hence get an estimate of the universal scaling function of equation~(\ref{pdsr}),
\begin{equation}
  \tilde{P}_L(x) = \frac{\hat{P}_L(Nx)}{\Delta X/\omega_X}.
  \label{eq:histo-est}
\end{equation}
Normalisation implies that
\[
\int P_L(x)\d x = \int bL^uP^\ast(bL^u\chi)\d\chi = \int P^\ast(\chi^\ast)\d\chi^\ast = 1,
\]
where $\chi^\ast \equiv b L^u\chi$. The corresponding distribution variances are
\begin{eqnarray}
  \sigma^2 &=& \int (x-x_0)^2P_L(x)\d x = \int bL^u\chi^2 P^\ast(bL^u\chi)\d \chi \nonumber \\
  & = & b^{-2}L^{-2u}\int {\chi^\ast}^2 P^\ast(\chi^\ast)\d \chi^\ast = b^{-2} L^{-2u}{\sigma^\ast}^2.
\end{eqnarray}
We can use the metric factors $b$ to ensure that the universal PDF $P^\ast$ has \emph{unit} variance, \emph{i.e.}, $\sigma^\ast = 1$. In that case, the basic relation (\ref{pdsr}) implies that
\begin{equation}
  P^{\ast}(\chi/\sigma)=\sigma P_L(x).
  \label{p*}
\end{equation}
We can then use the histogram (\ref{eq:histo-est}) to estimate the universal PDF $P^\ast$. The convenience of the above procedure is now evident: From standard Monte Carlo simulations we are able to estimate $P_L(x)$ as well as the standard deviation $\sigma$ and, without the need for knowing the exponent $u$, we can easily estimate the universal PDF $P^{\ast}$ from equation~(\ref{p*}).

While this formalism is fairly general, the most relevant cases clearly are $x = m$ for the magnetisation (or another order parameter such as the sub-lattice magnetisation $m_{j}$) and $x = e$ for the internal energy. For $x = m$ it has been shown in the original reference~\cite{binder} that $u=\beta/\nu$, while for $x=e$ one finds $u = (1-\alpha)/\nu$~\cite{milchev}. Here, $\alpha$ refers to the specific-heat exponent, while $\beta$ is the exponent of the magnetisation, and $\nu$ denotes the correlation-length exponent. Note that the exponent $u$ is just the scaling dimension of the given operator. We also note that for the case of $x = e$, equation~(\ref{dp}) is identical to the numerical estimate of the probability density function used for the EPD zeros approach discussed above. In addition to the data collected in that context, we found it worthwhile to conduct a single, long simulation at the best available estimate of the critical temperature, which is the basis for the results presented below. Finally, also histogram reweighting techniques may be used to improve statistics~\cite{alan}. Also, it is easily possible to work out the elements of $\Omega_X$. For the Baxter-Wu model at zero crystal field, for example, the energies lie in the symmetric range $-2L^2 S^3\le E\le 2L^2S^3$ with spacing $\Delta E = 1$ for both $S = 1$ and $S=3/2$.

\section{Monte Carlo simulations: setup and parameters}
\label{sec:simulations}

The Monte Carlo simulations reported in this paper were carried out using the single spin-flip Metropolis algorithm on square (Blume-Capel model) and triangular (Baxter-Wu model) lattices of linear size $L$ with periodic boundary conditions. Simulations were conducted for several values of the crystal field in the range $-2 \leq \Delta \leq 1$ in order to carefully investigate the question of universality in the two models. Owing to the uncertainty in the precise location of the pentacritical point as well as the observed strong corrections to finite-size scaling for the spin-$1$ Baxter-Wu model (see below), our simulations only reached up to $\Delta = 0.5$ for this case. In all simulations we set $J = 1$ and $k_{\rm B} = 1$, so that the crystal-field coupling and the temperature are measured in units of $J$ and $J/k_{\rm B}$, respectively. Note that for the Baxter-Wu model in order to properly accommodate the three different ferrimagnetic phases at low temperatures, all values of $L$ were selected as multiples of three. Further, the number of sweeps or Monte Carlo steps per spin (MCS) used for thermalisation and for computing average values of the thermodynamic quantities were chosen after several test runs designed to estimate the requirements for different system sizes $L$ as well as different values of the spin $S$ and the crystal field $\Delta$.

For the analysis of the EPD zeros, we considered lattices in the range $16 \le L \le 128$ for the Blume-Capel model and $18 \le L \le 120$ for the Baxter-Wu model, respectively. During thermalisation the first $N_{\rm therm}=10^5$ (resp.\ $3\times 10^5$) MCS were discarded for $L\le 45$ (resp.\ $L > 45$). The histograms were then obtained with a total of $N_{\rm MCS}=10^8$ MCS and the corresponding complex zeros were computed through the {\it GNU Scientific Library}~\cite{gnu}, which uses balanced-QR reduction of the companion matrix. The criterion to halt the iteration process of getting the pseudocritical temperatures $T_{\rm c}^L$ was considered to be satisfied when $T_0^{j+1}-T_0^j\le \eta$, with $\eta=10^{-4}$. Equivalent results were obtained when considering $|\operatorname{Re}(x_c^L)-1| \le \eta$. 

For both models, after obtaining the critical temperature, estimates of universal PDFs of the energy and magnetisation were usually computed using the larger system sizes with additional simulations comprising $N_{\rm MCS}=12\times 10^8$ MCS after thermalisation. When necessary, single histogram reweighing techniques were used to obtain the PDFs close to the estimated critical temperatures~\cite{alan}. 

\section{Results}
\label{sec:results}

\subsection{The Blume-Capel model}
\label{sec:bc_results}

We first applied the approaches presented above in sections~\ref{sec:theory} and \ref{sec:simulations} to the Blume-Capel model that is better understood than the Baxter-Wu model, thus creating a reference for the simulations of the latter, but also to gauge the reliability and accuracy of the methods.

\begin{figure}
\centering
\includegraphics[clip,width=0.8\hsize]{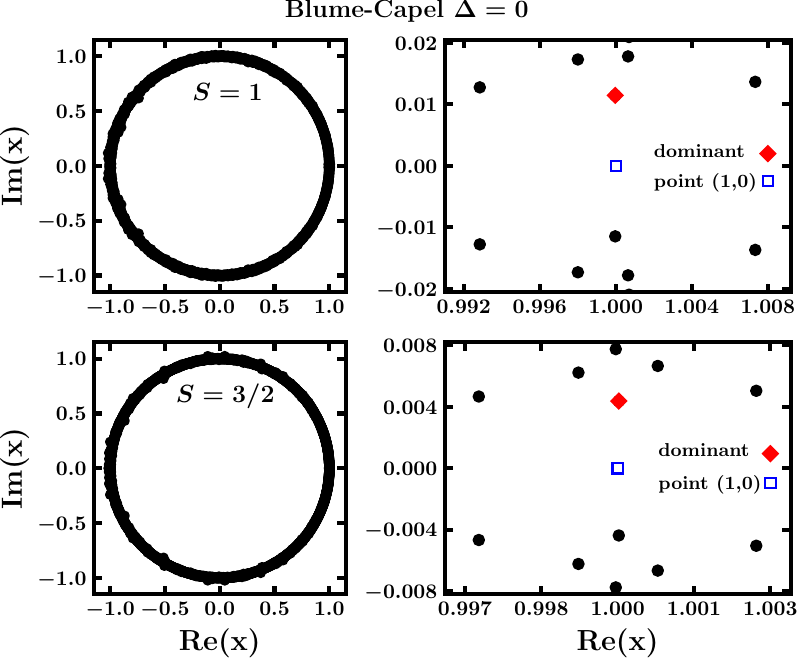}
\caption{Distribution of the EPD zeros in the complex plane for the Blume-Capel model at $\Delta=0$ and at $T_{\rm c}^L$. Results for $S=1$ (upper panels) with $L=80$ and $S=3/2$ (lower panels) with $L=100$ are shown. The panels on the right show in more detail the regions closer to the dominant zero. The diamond symbols mark the dominant root and the open squares are the reference point $(1,0)$, as indicated in both panels.}  
\label{zeros132d0}
\end{figure}

In figure~\ref{zeros132d0} we show the distribution of the EPD zeros in the complex plane for the Blume-Capel model at $\Delta=0$ and at the pseudo-critical temperatures $T_{\rm c}^L$. Results for both spins, \emph{i.e.}, $S=1$ and $S=3/2$, are shown. In particular, the right panels present a magnified view around the dominant root $x_{\rm c}^L$, illustrating that indeed the imaginary part $\operatorname{Im}(x_{\rm c}^L)$ is close to zero, while the real part $\operatorname{Re}(x_{\rm c}^L)$ is close to one. For each value of the spin $S$, the same pattern of the distribution of the EPD zeros is obtained, not only for different system sizes (with the number of roots rapidly increasing with $L$) but also for different values of $\Delta$.

\begin{figure}
\centering
\includegraphics[clip,width=0.7\hsize]{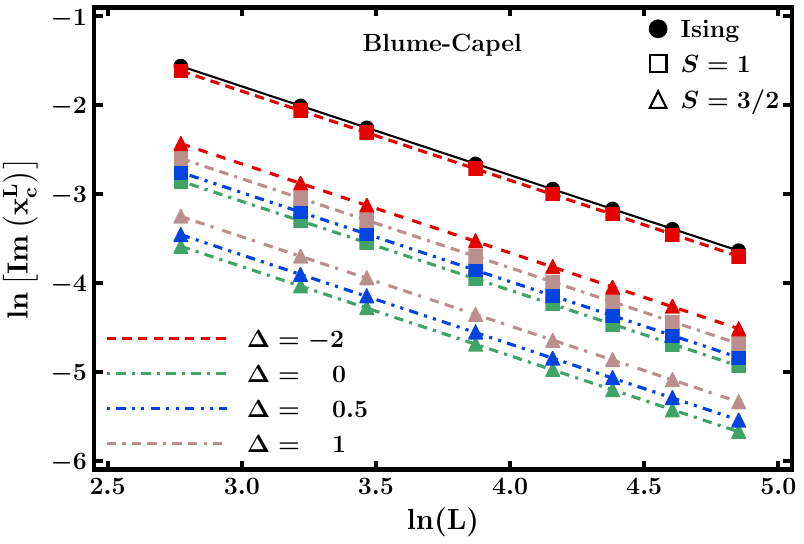}
\caption{Finite-size scaling behaviour of the imaginary part $\operatorname{Im}(x_{\rm c}^L)$ of the dominant EPD zero for the Blume-Capel model with spin $S=1$ (squares) and $S=3/2$ (triangles) at several values of the crystal-field coupling $\Delta$. For comparison, data for the spin $S=1/2$ case, \emph{i.e.}, the Ising ferromagnet, are also shown (circles). Note the doubly-logarithmic scale of the axes.}
\label{nubc}
\end{figure}

The finite-size scaling analysis of the imaginary part $\operatorname{Im}(x_{\rm c}^L)$ is depicted in figure~\ref{nubc} for the spin values $S=1$ and $S=3/2$ as well as for the full spectrum of $\Delta$ values considered in this work. As a reference, we also include results for the Ising model. All lines are linear fits to equation~(\ref{imx}) (excluding the correction term) with the magnitude of the slope providing $1/\nu$. The corresponding fit results are listed in table~\ref{tcnu}. It is interesting to note that all fitted lines are parallel to each other and to that of the Ising data, leading to the conclusion of a shared critical exponent $\nu = 1$ corresponding to the Ising universality class. This is verified to a good numerical accuracy by the extrapolated values reported in table~\ref{tcnu} for both values $S$ of the spin, as we move along the second-order transition line (always for $\Delta < \Delta_{\rm t} \approx 1.966$).

\begin{figure}
\centering
\includegraphics[clip,width=0.8\hsize]{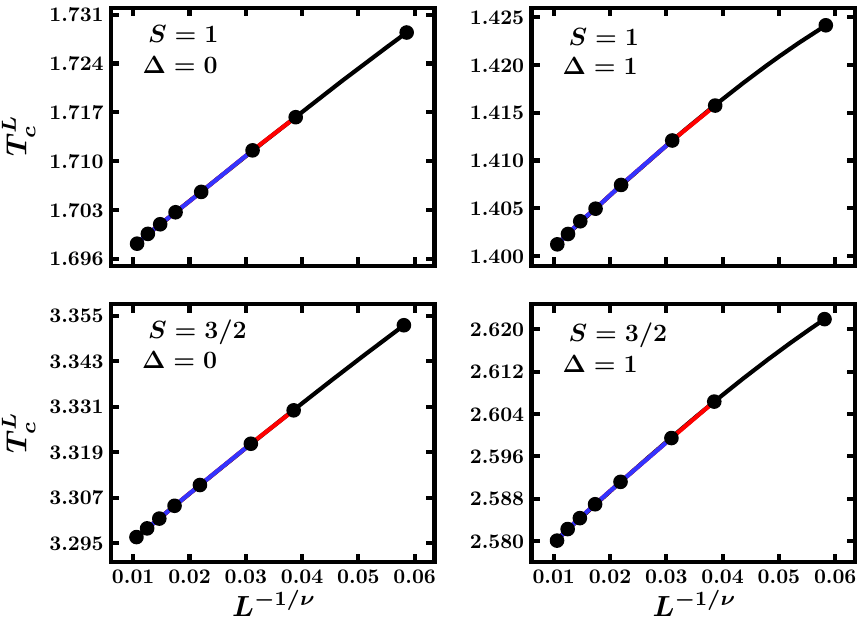}
\caption{Finite-size scaling behaviour of the pseudocritical temperatures $T_{\rm c}^L$ as extracted from the EPD zeros for the $S=1$ (upper panels) and $S=3/2$ (lower panels) Blume-Capel models at two values of the crystal field, $\Delta = 0$ (left panels) and $\Delta = 1$ (right panels). The lines show fits of the functional form (\ref{tc}) to the data (see main text for details).}
\label{tcbc}
\end{figure}

At this point we can use equation~(\ref{tc}) to estimate the critical temperatures $T_{\rm c}$. Figure~\ref{tcbc} presents the scaling behaviour of the pseudocritical temperatures $T_{\rm c}^{L}$ of the $S=1$ and $S=3/2$ Blume-Capel model at two values of the crystal field, as indicated. The numerical data are plotted against $L^{-1/\nu}$, with the ratio $1/\nu$ taken from table~\ref{tcnu}. Here, there is no significant change in the fits if we consider $L^{-(1/\nu) \pm \sigma(1/\nu)}$, where $\sigma(1/\nu)$ is the corresponding error in the critical exponent estimate. In each panel we show three separate fits: two linear fits [assuming $b^\prime = 0$ in equation~(\ref{tc})] with $L > 24$ (red line) and $L > 30$ (blue line), and an additional fit taking into account corrections to scaling, by fixing the exponent $\omega$ to the value $7/4$ found for magnetic quantities in the Ising model~\footnote{See, for example, the discussion in the supplementary material of reference~\cite{sandvik}. We note, however, that other values of $\omega$ have also been reported for certain quantities in the 2D Ising model, most notably $\omega = 4/3$ and $\omega = 2$, and in some cases also the analytic corrections might be dominant. In our case, the observed corrections are so weak (compared to the statistical accuracy of our data) that numerically we hardly see a difference between these choices.} (black line). Although for the case of $\Delta > 0$ ($\Delta = 1$ in this instance) the numerical data start to very slightly deviate from a straight line, all results for $T_{\rm c}$ are quite comparable and their average is shown in table~\ref{tcnu}.

We note that, for technical reasons, we did not systematically study the statistical fluctuations in the location of the leading zero\footnote{The results of reference~\cite{ronaldo} for the spin-$1/2$ Ising model showed that, although the overall map of zeros fluctuates substantially as a result of noise in the histogram, the location of the dominant zero is relatively stable.}. Instead, we merely estimate the statistical uncertainty in the estimates for $1/\nu$ and $T_c$ from the error estimates of fits of the functional forms (\ref{tc}), (\ref{eq:x_c}) and (\ref{imx}) to the data.  Note that in table~\ref{tcnu} we also include previous results from series expansion~\cite{butera} and conformal invariance~\cite{xav, dias17}, clearly suggesting an overall acceptable agreement regarding the critical temperatures of the Blume-Capel model along the second-order transition line.

\begin{table}[h!]
\caption{Summary of the main results obtained in the current work from the recently developed EPD zeros method~\cite{bis1,bis2,bis3} for the two-dimensional Blume-Capel and Baxter-Wu models. For  comparison the third column of the table includes some earlier reference estimates of critical temperatures ($T_{\rm c}^{\rm (ref)}$) obtained from series expansion~\cite{butera} and conformal invariance~\cite{xav,dias17}. The result $T_{\rm c}^{\rm (ref)} = 1.5300$ for the $S=1$ Baxter-Wu model at $\Delta = 0.5$ has been privately communicated to us by the authors of reference~\cite{dias17}. We note the expected values $1/\nu = 1$ and $1.5$ for the Blume-Capel and Baxter-Wu models, respectively.}
\begin{center}
\begin{tabular}{cccc}  
  \hline \hline
  \noalign{\vskip 1mm}
  ~~~$\Delta$~~~~~~~& ~~~~$T_{\rm c}$~~~~~ & ~~~ {$T_{\rm c}^{\rm (ref)}$}& ~~~~~~~{$1/\nu$}~~~~~~~  \\
  \noalign{\vskip 1mm}
  \hline \hline
   \noalign{\vskip 1.5mm}
  \multicolumn{3}{c}{Blume-Capel} & 1\\
  \noalign{\vskip 1.5mm}
  \hline \hline  
   \noalign{\vskip 1.5mm}
  \multicolumn{4}{c}{$S=1$} \\
  \noalign{\vskip 1.5mm}
  \hline
  \noalign{\vskip 1mm}	
   
   -2 ~~~~  &  ~~2.0013(3)~ &               & ~~1.001(2)  \\

   0 ~~~     &  ~~1.6938(1)~ & ~~~~~ 1.69378(4)~\cite{butera} & ~~1.001(2)     \\

   0.5~       &  ~~1.5662(1)~ & ~~~~~ 1.5664(1)~\cite{butera}   & ~~1.002(2)     \\

   1~~~~     &  ~~1.3977(1)~ & ~~~~~ 1.3986(1)~\cite{butera}  & ~~1.005(2)     \\
   
  \noalign{\vskip 1mm}
  \hline  
  \noalign{\vskip 1.5mm}
  \multicolumn{4}{c}{$S=3/2$} \\
  \noalign{\vskip 1.5mm}
  \hline
  \noalign{\vskip 1mm}
   -2 ~~~~  &  ~~4.1187(2)~ &             & ~~1.002(2)  \\

   0 ~~~~   &  ~~3.2884(6)~ &    ~~ 3.287(2)~\cite{xav}          & ~~1.004(2)  \\

   0.5 ~      &  ~~2.9727(3)~ &     ~~ 2.972(3)~\cite{xav}        & ~~1.003(2)   \\

   1 ~~~~   &  ~2.5740(3) &                                                                 & ~~1.004(2)     \\
   \hline \hline
  \noalign{\vskip 1.5mm}
  \multicolumn{3}{c}{Baxter-Wu} & 1.5 \\
  \noalign{\vskip 1.5mm}
  \hline \hline  
   \noalign{\vskip 1.5mm}
  \multicolumn{4}{c}{$S=1$} \\
  \noalign{\vskip 1.5mm}
  \hline
  \noalign{\vskip 1mm}	
   
   -2 ~~~~  &  ~~1.9796(4) &               & ~~1.503(1)   \\

   -1~~~~~ &  ~~1.8502(3) &  1.8503~\cite{dias17}~  & ~~1.515(2)   \\

   0 ~~~     &  ~~1.6606(5) &  1.6606~\cite{dias17}~  & ~~1.541(2)    \\

   0.5~       &  ~~1.5301(3) &  1.5300~~~   & ~~1.605(3)    \\
   
  \noalign{\vskip 1mm}
  \hline  
  \noalign{\vskip 1.5mm}
  \multicolumn{4}{c}{ $S=3/2$} \\
  \noalign{\vskip 1.5mm}
  \hline
  \noalign{\vskip 1mm}
   -2 ~~~~  &  ~~5.6645(5) &             & ~~1.523(2)    \\

   -1 ~~~~  &  ~~5.2576(2) &  5.2661~\cite{dias17}~ & ~~1.543(1)  \\
   
%   -0.5 ~~   &  ~~5.0037(5) &              & ~~1.571(1)    \\

   0 ~~~~   &  ~~4.7057(6) &              & ~~1.607(2)    \\

   0.5 ~      &  ~~4.3839(5) &              & ~~1.667(4)     \\
   
  \noalign{\vskip 1mm}
  \hline
  \hline
\end{tabular}\label{tcnu}
\end{center}
\end{table}

\begin{figure}
\centering
%\includegraphics[clip,width=0.7\hsize]{figures/Pebc-is-s132d051.pdf}
%\centering
\includegraphics[clip,width=0.7\hsize]{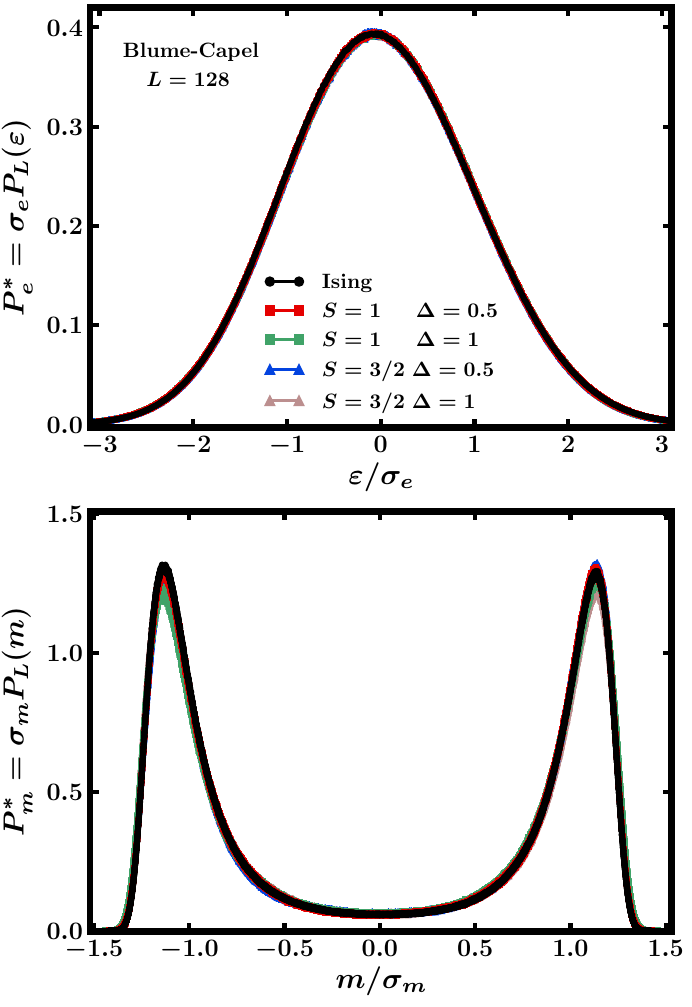}
\caption{Universal PDFs of the shifted energy density $\varepsilon = e-e_0$ (upper panel) and the magnetisation density $m$ (lower panel) in the Blume-Capel model for a system of linear size $L = 128$. Several distributions are illustrated for spin $S=1$ (squares) and $S=3/2$ (triangles) and two values of the crystal field, namely $\Delta=0.5$ and $\Delta=1$. For comparison, the reference distributions of the Ising ferromagnet (circles) are also sketched. Here, $\sigma_E$ and $\sigma_M$ denote the standard deviations of the energy and magnetisation histograms, respectively, see section~\ref{upd} for details.}  
\label{pstembc}
\end{figure}

While there is consistency in the value of $1/\nu$ along the transition line, cf.\ the data in table~\ref{tcnu}, further information is required to uniquely characterise a universality class. Here, we turn our attention to the universal PDFs of the main thermodynamic observables, in particular the energy and magnetisation, to achieve a fuller characterisation. In figure~\ref{pstembc} we show our estimates of the energy and magnetisation PDFs for the Blume-Capel model with spins $S=1$ and $S=3/2$ and crystal fields $\Delta=0.5$ and $\Delta=1$, respectively. These distributions were computed for the lattice size $L = 128$ at the corresponding pseudocritical temperatures of the system. We underline that on the scale of figure~\ref{pstembc}, the results for $L = 80$ and $L = 100$ (not shown) fall on top of those for $L=128$, in particular for the energy distribution. For reference, also the PDFs for the spin $S=1/2$ Ising ferromagnet are included in both panels of figure~\ref{pstembc}.

Without doubt, the energy PDFs shown in figure~\ref{pstembc} constitute a strong indication of Ising universality. Although the same occurs for the magnetisation distribution functions, some variations can be observed around the two symmetric peaks. These might be due to increased fluctuations implied by critical slowing down in the Metropolis (single-spin flip) dynamics. This problem could certainly be alleviated by using hybrid algorithms such as single-spin flips combined with Wolff cluster updates, a practice that was shown to effectively improve the simulations in the Blume-Capel model~\cite{hybrid,fytas18}. However, we decided to implement in this work only the Metropolis algorithm for reasons of consistency with the parallel study of the Baxter-Wu model, for which an effective hybrid procedure is harder to establish.

In summary, with the help of only the Metropolis algorithm and investing a rather moderate computational effort, the energy PDF proves to be a robust tool in ascertaining the critical behaviour and universality of the Blume-Capel model. This suggests that the energy PDF could be an underestimated tool in numerical studies of critical phenomena. 

\subsection{The Baxter-Wu model}
\label{sec:bw_results}

In a previous work~\cite{arilton}, we already studied the $S=1$ Baxter-Wu model with the help of the EPD zeros method, and we will use some of the previously obtained numerical estimates to facilitate the discussion. A summary of results for both the $S=1$ and $S=3/2$ models is shown in table~\ref{tcnu}, together with previous estimates from conformal invariance~\cite{dias17}. 

\begin{figure}
\centering
\includegraphics[clip,width=0.7\hsize]{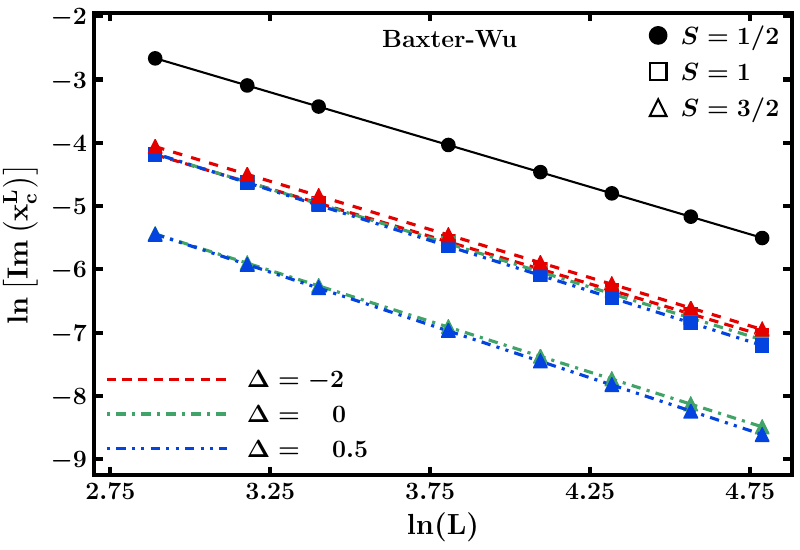}
\caption{Finite-size scaling behaviour of the imaginary part $\operatorname{Im}(x_{\rm c}^L)$ for the Baxter-Wu model with spin $S=1$ (squares) and $S=3/2$ (triangles) at several values of the crystal-field coupling $\Delta$. The circles in this case refer to the spin $S=1/2$ model.}  
\label{nubw}
\end{figure} 

When we apply the EPD zeros method to the $S=3/2$ model, the distribution of zeros in the complex plane and the corresponding scaling plots of $\operatorname{Im}(x_{\rm c}^L)$ and $T_{\rm c}^L$ (for the latter using now the putative correction term with $\omega = 2$) all appear very similar to those for the $S=1$ case (not shown), with the exception of the asymptotic values of non-universal quantities such as the transition temperatures. Comparing to previous results, we find that the critical-temperature estimates from the EPD zeros method are comparable to those from conformal invariance within error bars for the $S=1$ model, while the agreement for $S=3/2$ is a bit less convincing, cf.\ the data collected in table~\ref{tcnu}. Increasing the number of spin states seems to require longer simulations.\footnote{Since the transition temperatures for this model were obtained considering $\eta=10^{-4}$, errors have been estimated by $(2-3)\sigma_f$, with $\sigma_f \sim 10^{-5}$ the variance of the corresponding fits.}

On the other hand, by renormalisation-group arguments the critical exponent $\nu$ is expected to maintain its original value of $2/3$ (or $1/\nu = 3/2$ in the notation used in the present work), independent of $S$ and $\Delta$, as long as we move along the second-order transition line of the phase boundary. A typical illustration that combines data for all spin values $S$ studied and various values of $\Delta$ is given in figure~\ref{nubw}, which is the analogue of figure~\ref{nubc} for the Baxter-Wu model. For comparison, results for the spin $S=1/2$ model are also shown. In contrast to figure~\ref{nubc}, the fits for the Baxter-Wu model appear to show slight deviations from the straight line as both $S$ and $\Delta$ increase and, in particular, for $\Delta \gtrsim 0$, which is also evident from the actual extrapolated $1/\nu$ values recorded in table~\ref{tcnu}. A similar behaviour was also observed for the $S=1$ model at $\Delta = 0$ in reference~\cite{arilton}, where it was attributed to the presence of strong finite-size effects due to the proximity to the putative multicritical point. Preliminary simulations of hybrid type consisting of suitable cluster updates~\cite{swendsen87,novotny93} with the heat-bath algorithm~\cite{miyatake86,loison04} showed that the critical exponent $\nu$ approaches the expected result when considering very large system sizes~\cite{arilton}. However, we should note that for negatives values of $\Delta$ a good agreement with the $S=1/2$ model is achieved, both for $S=1$ and for $S=3/2$, as is evident from table~\ref{tcnu}.

In order to resolve this conundrum, it is worthwhile to explore in detail the universal PDFs of the energy and magnetisation of the Baxter-Wu model, a task involving considerably less computational effort than the high-precision studies of the critical exponents. To set the stage, we first consider these distributions for the case $S=1/2$, where there is some previous work for the energy~\cite{adler,velo} and total magnetisation~\cite{marti,velo0}. Here, we provide more accurate data for larger systems, and we also include an analysis based on the sublattice magnetisations~\cite{vasilopoulos22, arilton}.

\begin{figure}
\centering
\includegraphics[clip,width=0.7\hsize]{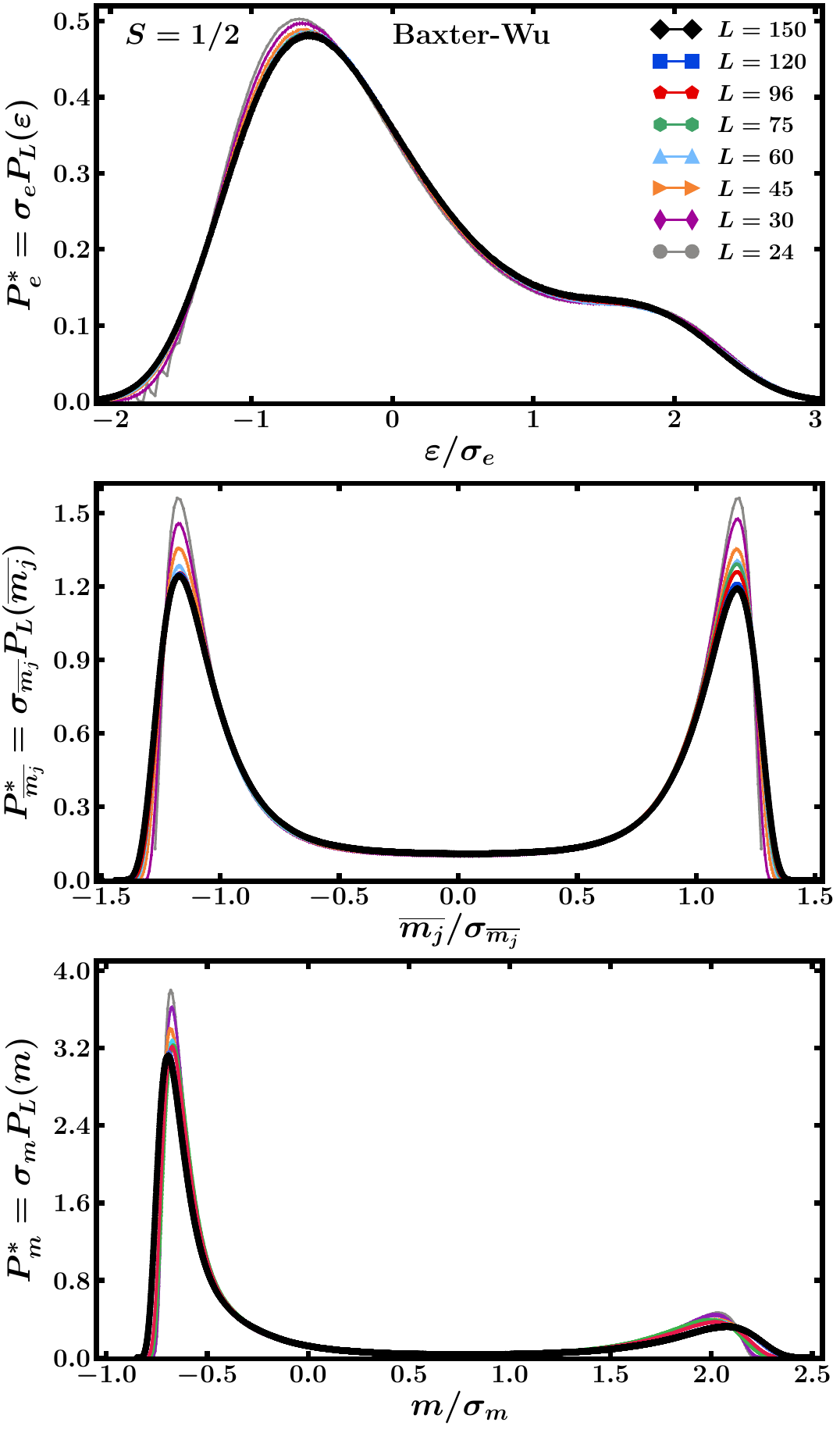}
\caption{Energy and magnetisation PDFs of the spin-$1/2$ Baxter-Wu model at the exact critical temperature for different lattice sizes $L$, as indicated. In particular: The upper panel illustrates the energy PDF ($P^{\ast}_e$), the middle panel that of the average over the three sublattice magnetisations ($P^{\ast}_{\overline{m_{j}}}$ with $j=1,2,3$), and the lower panel that of the total magnetisation ($P^{\ast}_m$).}  
\label{pdfs05}
\end{figure}

The corresponding PDFs of the pure spin $S = 1/2$ Baxter-Wu model, computed at the exact critical temperature $T_{\rm c} = 2.26918\cdots$, are shown in figure~\ref{pdfs05} for several lattice sizes. On the scale of the graph, the energy universal PDF is achieved for $L \ge 60$, while for the magnetisation, a  universal PDF is achieved only for the larger lattices $L \ge 120$. While the total magnetisation has a higher peak for a negative value of $m$ (reflecting the fact that two out of the three sublattices have negative spin orientation), the sublattice magnetisations do show a symmetric distribution~\cite{comment1}.
\begin{figure}
\centering
\includegraphics[clip,width=0.7\hsize]{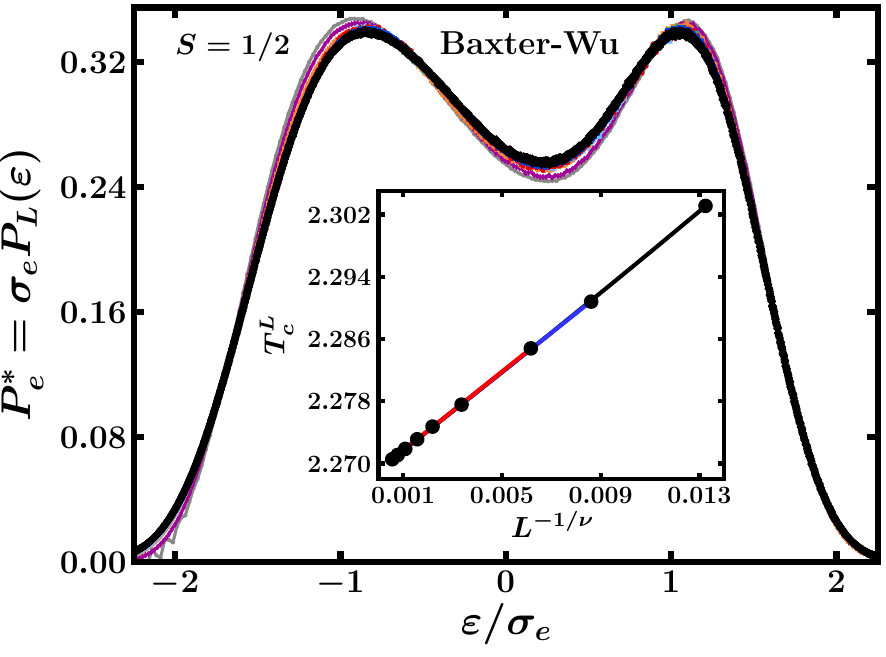}
\caption{Energy PDFs of the spin-$1/2$ Baxter-Wu model for the same sequence of lattice sizes as in figure~\ref{pdfs05}. The PDFs were computed at the temperatures where the distributions show two peaks of equal height. The inset shows the finite-size scaling behaviour of the equal-height pseudocritical temperatures $T_{\rm c}^L$, where the different coloured lines correspond to the three fits corresponding to figure~\ref{tcbc} that were outlined in the previous section.} 
\label{P*difT}
\end{figure}

It is worth noting that for a temperature just above the critical one, $T = 2.27055 > T_{\rm c} = 2.26918\cdots$ and system size $L = 150$, the right shoulder in the energy PDF shown in the upper panel of figure~\ref{pdfs05} evolves into a second peak of the same height as the left one (not shown). This double-peaked function in the energy distribution has been previously interpreted as a sign of a first-order transition in the model~\cite{lucas1st}. Using the histogram reweighting~\cite{alan} we show in figure~\ref{P*difT} for the system sizes studied the energy PDFs computed at the (system-size dependent) temperature where the two peaks are of equal height. There is a clear agreement with the previous results of figure~\ref{pdfs05}, as also here we document graphically the convergence towards a unique density function for $L \ge 60$. This establishes the equal-height temperature as a new pseudocritical temperature of the system. Fitting the functional form of equation~(\ref{tc}) to this sequence of pseudocritical points one arrives at the estimate $T_{\rm c} = 2.2692(1)$~\cite{comment2}, in excellent agreement with the exact result.  

Reviewing this first part of results for the spin-$1/2$ Baxter-Wu model, we should emphasise that figure~\ref{P*difT} represents an extension to the energy PDF of a method originally proposed for the magnetisation~\cite{puli} in determining the critical temperature when one does not know, a priori, the universal function. Despite the presence of the double peaks, the analysis in figure~\ref{P*difT} also confirms the expected second-order character of the transition, corroborating recent results for the spin-$1$ model based on a scaling analysis of the surface tension and latent heat at $\Delta \leq 0.5$~\cite{vasilopoulos22,arilton}.

\begin{figure}
\centering
\includegraphics[clip,width=0.7\hsize]{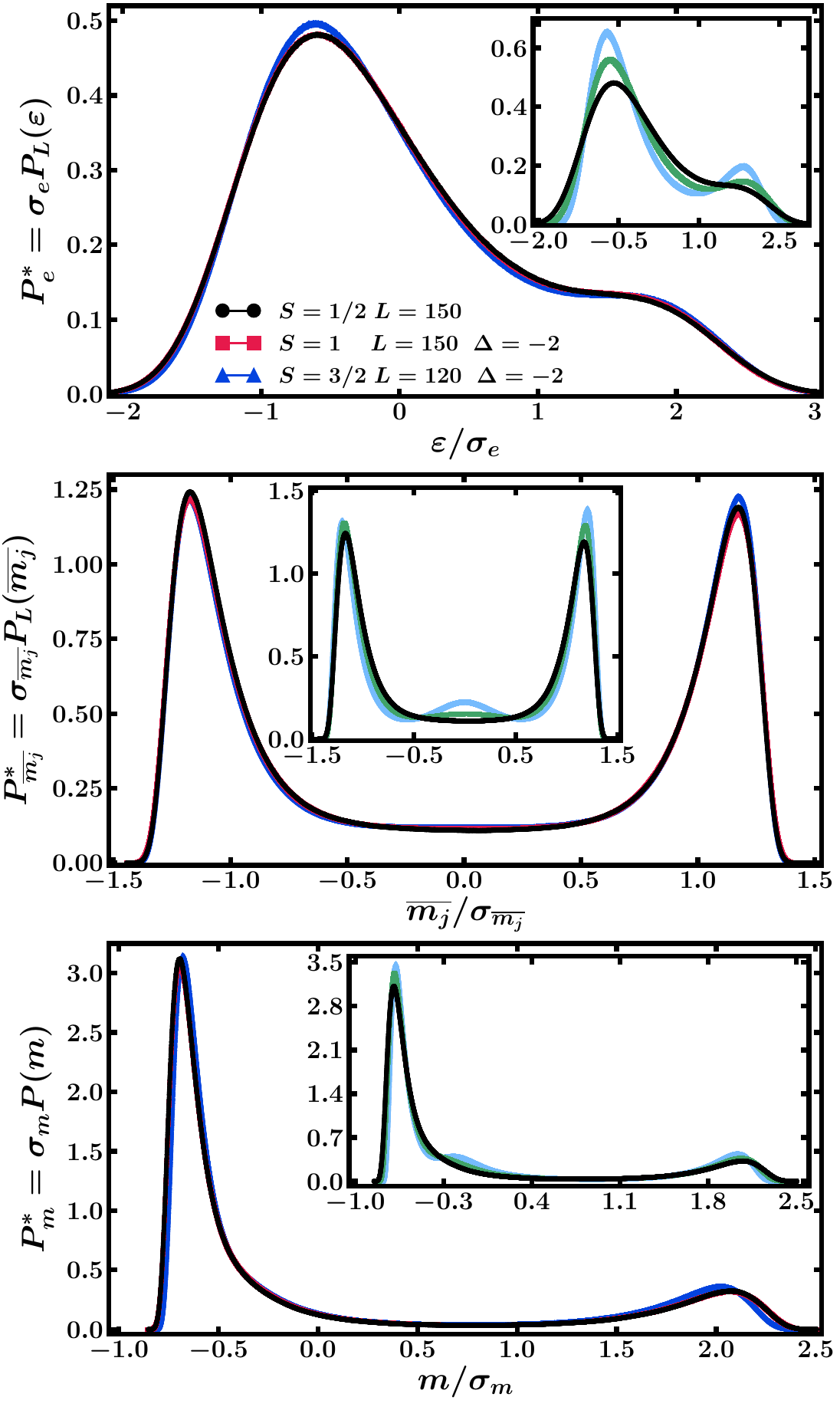}
\caption{Energy and magnetisation PDFs of the Baxter-Wu model with $S=1$ (squares) and $S=3/2$ (triangles) at two values of the crystal field, namely $\Delta=-2$ (main panels) and $\Delta=0.5$ (insets: green symbols for $S=1$ and blue symbols for $S=3/2$, respectively). The upper panel illustrates the energy PDF ($P^{\ast}_E$), the middle panel that of the average over the three sublattice magnetisations ($P^{\ast}_{\overline{m_{j}}}$ with $j=1,2,3$), and the lower panel that of the total magnetisation ($P^{\ast}_m$). The universal PDFs of the spin-$1/2$ model (circles) are also plotted for reference.}  
\label{pstembw}
\end{figure}

\begin{figure}
\centering
\includegraphics[clip,width=0.7\hsize]{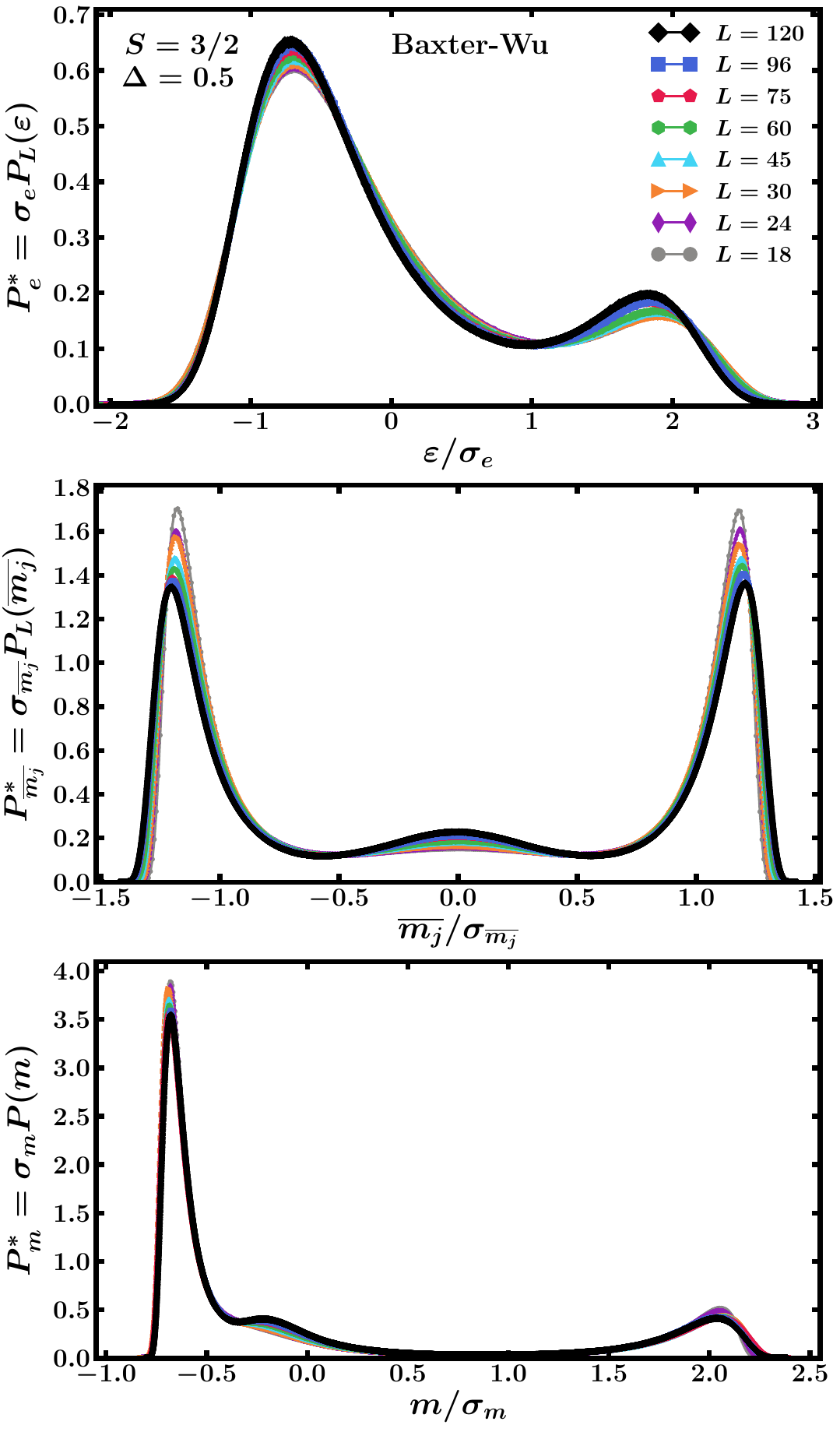}
\caption{Energy and magnetisation PDFs of the Baxter-Wu model with $S=3/2$ and $\Delta=0.5$ for several values of the lattice size $L$. The upper panel illustrates the energy PDF ($P^{\ast}_e$), the middle panel that of the average over the three sublattice magnetisations ($P^{\ast}_{\overline{m_{j}}}$ with $j=1,2,3$), and the lower panel that of the total magnetisation ($P^{\ast}_m$).}  
\label{pdfbws32d05}
\end{figure}

We now turn to the PDFs of the spin-$1$ and spin-$3/2$ Baxter-Wu models. These particular PDFs have been computed with much longer simulation times~\cite{comment3} at the estimated critical temperatures as listed in table~\ref{tcnu}; these PDFs are shown in figure~\ref{pstembw}. For comparison, we also show the PDFs of the spin-$1/2$ model. While for $\Delta \lesssim 0$ and smaller $S$ the PDFs of different models collapse onto the same universal functions, we observe some deviations in particular for $S=3/2$ and $\Delta = 0.5$. In order to gauge these, we show in figure~\ref{pdfbws32d05} the system-size dependence of the observed PDFs, which are observed to  be much more pronounced here than for the pure $S=1/2$ Baxter-Wu model, cf.\ figure~\ref{pdfs05}. Nevertheless, it appears that for the largest system sizes considered, the PDFs have already stabilised to a certain degree, but this impression might be deceptive.

Inspecting the results of figure~\ref{pstembw}, we hence come to the following conclusions: (i) For $\Delta=-2$ and both values of the spin, all three PDFs manifest a reasonable agreement with the universal ones coming from the spin-$1/2$ case. We note that this trend is more apparent for the magnetisation PDFs and becomes even more definite for crystal-field values $\Delta < -2$. (ii) For positive values of $\Delta$ (as shown in the insets of figure~\ref{pstembw}) the situation appears to be much more involved. For example, at $\Delta=0.5$ the distributions of both spin $S=1$ and $S=3/2$ models appear to show deviations from that observed for the spin-$1/2$ model. In fact, the energy PDF starts to develop a secondary peak, where initially one has a shoulder and the sublattice magnetisation PDF presents an additional peak at zero magnetisation ($m_{j} = 0$). We attribute these discrepancies among the PDFs, which appear to become more pronounced upon increasing the crystal field $\Delta$ in the direction of the pentacritical point and the spin $S$ value, to the same finite-size effects that obscured the analysis of the EPD zeros method in the previous section.
 
\section{Concluding remarks}
\label{sec:summary}

In the present paper we analysed several critical aspects of the two-dimensional Blume-Capel and Baxter-Wu models in the presence of a crystal-field coupling $\Delta$ and for various values of the spin $S$. We benefited from a recently proposed method utilising the zeros of the energy probability distribution as well as from the physical information encoded in the universal probability distribution functions of the energy and magnetisation. Numerically, we employed extensive Monte Carlo simulations based on the Metropolis algorithm in combination with single-histogram techniques.

For the Blume-Capel ferromagnet, the reported original results are in excellent agreement with the expected behaviour. Namely, our estimates for the critical exponent $\nu$ of the correlation length are fully consistent with the Ising universality class, a result which is further reinforced by the considered probability density functions of both the energy and magnetisation. Additionally, the critical temperatures $T_{\rm c}(\Delta)$ obtained from standard finite-size scaling are comparable to some of the best known estimates from the recent literature.
Similar conclusions in general apply also for the Baxter-Wu model, where both the computation of the critical exponent $\nu$ but also the universal shape of the probability density functions suggest that all studied spin-$S$ models share the universality class of the $4$-state Potts model. We remind the reader that this in principle anticipated from symmetry arguments~\cite{domany78} and is also in agreement with recent high-accuracy numerical results for the spin-$1$ model~\cite{vasilopoulos22}. Still, an intriguing observation emerging from our simulations is the slight deviation of the exponent $\nu$ from the expected $2/3$ result as well as various mismatches in the probability density functions upon increasing $\Delta$ and $S$, most strongly visible for $S=3/2$ and $\Delta = 0.5$. Apparently, the problem becomes much more involved for positive values of $\Delta$, requiring simulations of much larger system sizes, a task which goes beyond the scope of the present work. 
In this regime, there appear to be strong finite-size effects that were also observed in previous studies of the model~\cite{vasilopoulos22,arilton}. 

One possible explanation for these deviations arises from the concept of \emph{field mixing}~\cite{wild1}. In studying first-order phase transitions close to a second-order line, with an intervening multicritical point, a mixing of scaling fields (resp.\ a demixing) turns out to be of paramount importance for identifying a suitable \emph{direction} that minimises corrections to scaling. As the crystal field increases, the second-order transition line gets steeper, bringing about a higher degree of asymmetry in the thermodynamic fields. In this respect, the process of a mixing of such thermodynamic fields may also be relevant along the tetracritical line. Hence, taking such effects into account might be crucial in order to accurately obtain the universal probability distribution functions. This aspect is a worthy subject for future investigations.

\section*{Acknowledgements}
We would like to thank Prof.\ Lucas M\'ol for fruitful discussions on the use of the EPD zeros method and Prof.\ Gerald Weber for invaluable assistance in the use of the Statistical Mechanics Computer Lab facilities at the Universidade Federal de Minas Gerais. The work of A. Vasilopoulos and N.~G. Fytas was supported by the  Engineering and Physical Sciences Research Council (grant EP/X026116/1 is acknowledged). This research was supported by CNPq, CAPES, and FAPEMIG (Brazilian agencies).

{}


\begin{thebibliography}{}
\bibitem{comment1} As a comment we note that in  this particular case of the $S=1/2$ Baxter-Wu model we extended our simulations to the size $L=180$. However, we observed that while the PDFs for the energy and total magnetisation are almost identical to those of $L=150$, the sublattice magnetisations have peaks of slightly different heights that alternate depending on the number of Monte Carlo steps (typical of the single spin-flip nature of the Metropolis algorithm). It appears possible that with increasing $L$, the system spends more time in one of the sublattice phases.
\bibitem{comment3} Except for the $S=1/2$ and $S=1$ models at $\Delta=-2$ where we used as a reference the system with linear size $L = 150$, for all the other cases we restricted our analysis to the size $L=120$. This is due to the fact that upon increasing $S$ and $\Delta$ simultaneously the necessary computational time to obtain a reasonable distribution becomes prohibitive.
\end{thebibliography}
\end{document}